\documentclass[10pt,a4paper]{article}
\usepackage{amsmath,amsthm,amssymb,amsfonts,fullpage, array,hyperref,longtable,cancel,mathtools}
\usepackage[utf8]{inputenc}
\numberwithin{equation}{section}

\newcommand{\Res}{\mathop{\mathrm{Res}}\limits}
\newtheorem{proposition}{Proposition}
\allowdisplaybreaks
\makeatletter
\def\section{\@startsection{section}{1}{\z@}%
            {-3.5ex \@plus -1ex \@minus -.2ex}%
            {2.3ex \@plus.2ex}%
            {\normalfont\large\bfseries}}
\def\subsection{\@startsection{subsection}{2}{\z@}%
            {-3.25ex\@plus -1ex \@minus -.2ex}%
            {1.5ex \@plus .2ex}%
            {\normalfont\normalsize \bfseries}}
\def\@seccntformat#1{\csname the#1\endcsname.~~}
\def\Appendix{\appendix
  \def\@seccntformat##1{Appendix~\csname the##1\endcsname.~~}}
\makeatother

\begin{document}
\newpage
\title{\Large Form factors in the Bullough-Dodd related models:\\ The Ising model in a magnetic field.}
\author{Oleg Alekseev\\[\medskipamount]
{\normalsize\it
Landau Institute for Theoretical Physics,}\\
{\normalsize\it 142432 Chernogolovka of Moscow Region, Russia}}
\date{}
\maketitle

\begin{abstract}
We consider particular modification of the free-field representation of the form factors in the Bullough-Dodd model. The two-particles minimal form factors are excluded from the construction.  As a consequence, we obtain convenient representation for the multi-particle form factors, establish recurrence relations between them and study their properties. The proposed construction is used to obtain the free-field representation of the lightest particles form factors in the $\Phi_{1,2}$ perturbed minimal models. As a significant example we consider the Ising model in a magnetic field. We check that the results obtained in the framework of the proposed free-field representation are in agreement with the corresponding results obtained by solving the bootstrap equations.
\end{abstract}

\section{Introduction}
As it is known two-dimensional statistical models in their critical points are described by the so called minimal models of the conformal field theory~\cite{BPZ}. Away from critical points the scaling region can be described by relevant perturbations of the fixed point action. The corresponding models can be referred to as the perturbed minimal models. The perturbations destroy long-range correlations of the critical model and the associated quantum field theories are usually massive. However in some cases an infinite number of integrals of motion survive. In particular, the minimal models perturbed by the one of the following primary operators, $\Phi_{1,2}$, $\Phi_{2,1}$, $\Phi_{1,3}$, are known to be integrable~\cite{ABZIMMF,ABZIMMF2}. Besides, in~\cite{T} it is shown that the $\Phi_{1,5}$ perturbation of non-unitary minimal models is also integrable. The $\Phi_{1,3}$ perturbed minimal models can be described as certain restrictions of the sine-Gordon model~\cite{RS,BerLe}. The other cases are related with particular quantum group restrictions of the Zhiber-Mikhailov-Shabat model~\cite{Smirnov,Eft,T}. 

The off-shell behavior of these models can be studied in the framework of the form factor approach~\cite{BKW}. In particular, correlation functions can be reconstructed using the spectral density representation. The fast rate of convergence of the spectral series for all distance scale~\cite{CM} allows one to calculate them quite accurately ignoring the multi-particle form factors. Besides, in the recent years another application of the form factor approach is deeply developing in the framework of studying a specific class of the two-dimensional non-integrable field theories. The non-integrable field theories can be treated as a particular perturbations of the integrable ones~\cite{DMS} and multi-particle form factors are significant in the study of these models. A particular example is provided by the Ising model whose off-critical behavior can be considered by using the two different integrable models. The recent progress in the studying of the magnetic deformation of the Ising model at non-critical temperature can be found in~\cite{ZZ}. The perturbation theory near another integrable point is more complicated~\cite{DelfinoM}. Some numerical data were obtained in~\cite{T1,T2} by means of the truncated conformal space approach.

The mentioned relation between different integrable models provides an effective method of the form factors calculation in the perturbed minimal models. For example, the $S$-matrix of the particles in the $\Phi_{1,3}$ perturbed minimal model originates from the breathers $S$-matrix of the corresponding sine-Gordon model whose off-shell behavior was considered in a number of papers~\cite{Mussardo,Mussardo2}. However for the $\Phi_{1,2}$, $\Phi_{2,1}$ and $\Phi_{1,5}$ perturbed minimal models the situation is quite different because of complexity of the $S$-matrix of the Zhiber-Mikhailov-Shabat model~\cite{Smirnov,Eft}. In this paper we consider a very particular class of the form factors which can be obtained from the form factors of the Bullough-Dodd model~\cite{BD,FMS,A}. The convenient method of the multi-particle form factor calculation is provided in the framework of the free-field representation~\cite{Lukyanov}. This method was successfully applied for finding multi-particle form factors in many integrable models~\cite{Lukyanov1,Lukyanov2,FateevL,FateevPP,FateevP}. The free-field representation  for the Bullough-Dodd model was proposed in~\cite{Shiraishi,BL}. However in some cases particular modification of the free-field representation is more convenient. In this representation the two-particle minimal form-factors are excluded from the construction. In~\cite{FL,AL} this method was applied for finding form factors of descendant operators. In this paper we consider another its feature. Namely, this representation possesses simple analytical properties and can be used to obtain convenient free-field representation for the lightest particles form factors in the $\Phi_{1,2}$ perturbed minimal models.

As it is known in the free-field representation the calculation of the multi-particle form factors reduces to a combinatoric procedure. The $N$-particle form factors in the Bullough-Dodd model and, consequently, in the corresponding perturbed minimal models are given by the sum of $3^N$ terms. In opposite to conventional construction in the proposed free-field representation Wick's averaging procedure generates functions with simple analytic properties. This feature allows us to obtain particular recurrence relations between the multi-particle form factors and prove the reflection properties for the form factors of the exponential operators that was suggested in~\cite{FLZZ}. Also, we prove that the form factors satisfy the quantum equation of motion. The certain restrictions of the imaginary coupling Bullough-Dodd model correspond to $\Phi_{1,2}$, $\Phi_{1,5}$ or $\Phi_{2,1}$ perturbations of the $\mathcal M_{p,p'}$ minimal models~\cite{Smirnov,Eft,T}. In this paper we consider the first possibility. The $S$-matrix of the lightest particles in these models~\cite{Koubek} correspond the $S$-matrix of the Bullough-Dodd model being analytically continued to the imaginary values of the coupling constant. This allows us to propose the free-field representation for the lightest particles form factors in the $\Phi_{1,2}$ perturbed minimal models. 

As a significant example we consider the Ising model in a magnetic field (IMMF). This model correspond to the $\Phi_{1,2}$ perturbation of the $\mathcal M_{3,4}$ minimal model and is known to be related with the $E_8$ algebra~\cite{ABZIMMF2}. It is a challenging problem to obtain the free-field representations associated with this algebra directly. The remarkable connection between IMMF and the Bullough-Dodd model was established in~\cite{Shiraishi,BL}. We present the free-field representation for the lightest particle of the IMMF explicitly and compare the calculations performed in the framework of the free-field representation with the corresponding results obtained in~\cite{DelfinoM,DM1} by solving the bootstrap equations. We checked that these results are in full agreement.

This paper is organized as follows. In section~\ref{ST} we review the general features of two-dimensional integrable quantum field theories and specify the notation we use. In section~\ref{BD} we present the modification of the free-field representation by excluding the two-particle minimal form factors from the construction. We discuss the features of the proposed representation and obtain particular recurrence relations between the form factors. By using this relations we prove the reflection properties  of the form factors of the exponential operators and show that these form factors satisfy the quantum equation of motion. In section~\ref{cBD} we consider quantum group restrictions of the complex Bullough-Dodd model which correspond to the $\Phi_{1,2}$ perturbations of the $\mathcal M_{p,p'}$ minimal models and propose the free-field representation for the lightest particles form factors in these models. As an example of the proposed construction the Ising model in a magnetic field is considered in detail.

\section{Scattering theory}\label{ST}
Let us consider the relativistic scattering theory containing $n$ sort of the particles $A_a$, $a=1,\ldots,n$ with masses $m_a$. We introduce the notation $A_a(\theta)$ for the particle $A_a$ having the rapidity $\theta$,
\begin{equation}\label{ST:rapidity}
	p_a^0=m_a\cosh\theta_a,\quad p_a^1=m_a\sinh\theta_a,
\end{equation}
where $p_a^0$, $p_a^1$ are components of the two-dimensional momentum $p_a$ such that the mass-shell condition is satisfied, i.e.\ $p_a^2=m_a^2$. The states
\begin{equation}\label{ST:AStates}
	|A_{a_1}(\theta_1),\ldots A_{a_n}(\theta_n)\rangle_{in(out)}
\end{equation}
form a basis of the asymptotic in(out)-states which is assumed to be complete. The rapidities of the states are supposed to be ordered, that is, $\theta_1>\ldots>\theta_n$ for incoming asymptotic states and $\theta_1<\ldots<\theta_n$ for outgoing ones.

The $S$-matrix is defined as the unitary transformation connecting ``in'' and ``out'' asymptotic states. Let us assume that all the particles $A_a$ have different masses. This assumption, in particular, implies that all the particles are neutral, $\bar A_a=A_a$, where $\bar A_a$ denotes the corresponding anti-particle. Let us restrict ourself to the case of the diagonal scattering only. Then the $S$-matrix is diagonal in the basis~\eqref{ST:AStates}
\begin{equation}\label{ST:S}
	|A_{a_1}(\theta_1),\ldots A_{a_n}(\theta_n)\rangle_{in}=S_{a_1 \ldots a_n}(\theta_1,\ldots,\theta_n)|A_{a_1}(\theta_1),\ldots A_{a_n}(\theta_n)\rangle_{out},
\end{equation}
An infinite number of the conserved currents in the integrable theories forces the scattering processes to be completely elastic, i.e.\ the final state contains the same number of particles with the same momenta as the initial one. Therefore, the $n$-particle $S$-matrix element in~\eqref{ST:S} factorizes in the product of $n(n-1)/2$ two-particle ones,
\begin{equation}\label{ST:S2}
	S_{a_1\ldots a_n}(\theta_1,\ldots,\theta_n)=\prod_{i<j}S_{a_i a_j}(\theta_i,\theta_j).
\end{equation}
The amplitudes $S_{ab}(\theta_1,\theta_2)$ are meromorphic functions in the rapidity difference $\theta=\theta_{12}=\theta_1-\theta_2$, real at ${\rm Re}\ \theta=0$ and satisfying the equations
\begin{equation}\label{ST:SS}
	S_{ab}(\theta)=S_{ab}(i\pi-\theta),\quad S_{ab}(\theta) S_{ab}(-\theta)=1,
\end{equation}
which express the crossing symmetry and the unitary conditions of the theory respectively. From these conditions it follows that the amplitudes $S_{ab}(\theta)$ are $2\pi i$-periodic functions, which are completely determined by positions of their zeros and poles in the ``physical strip'' $0\leq {\rm Im}\ \theta\leq\pi$. The poles on the strip are located at ${\rm Re}\ \theta=0$. Simple poles with positive residue correspond to ``bound state'' particles in the $s$-channel of the $A_a A_b$ scattering, whereas those with negative one correspond to ``bound states'' in the $u$-channel. The basic bootstrap requirement is that the ``bound states'' must belong to the same set of the particles $A_a$, $a=1,\ldots,n$. The simple poles with positive residue of the two-particle amplitudes $S_{ab}(\theta)$  at $\theta=iu_{ab}^c$ represents the particle $A_{c}$ with mass
\begin{equation}\label{ST:u}
	m_c^2=m_a^2+m_b^2+2m_am_b\cos u_{ab}^c,\quad u_{ab}^c\in(0,\pi),
\end{equation}
in the $s$-channel of the $A_a A_b$ scattering. In the vicinity of the bound state pole the two-particle amplitude becomes
\begin{equation}\label{ST:Sresu}
	S_{ab}(\theta\simeq iu_{ab}^c)\simeq\frac{i(\Gamma_{ab}^c)^2}{\theta-iu_{ab}^c},
\end{equation}
where $\Gamma_{ab}^c$ can be referred to as the on-shell three-particle coupling constant.

The off-shell behavior of the integrable models can be studied in the form factor framework approach. For the local operator $\mathcal O(x)$ its form factors are defined as the matrix elements of this operator in the basis of the asymptotic states,
\begin{equation}\label{ST:FF}
	F^{\mathcal O}_{a'_1\ldots a'_m;a_1 \ldots a_n}(\theta'_m, \ldots,\theta'_1|\theta_1,\ldots,\theta_n)\ =\ _{a'_1 \ldots a'_m}\langle\theta'_1,\ldots,\theta'_m|\mathcal O(0)|\theta_1,\ldots,\theta_n\rangle_{a_1 \ldots a_n}.
\end{equation}
The crossing symmetry condition of the $S$-matrix~\eqref{ST:SS} suggest that the form factors can be expressed in terms of the following matrix elements,
\begin{equation}\label{ST:elFF}
	F^{\mathcal O}_{a_1 \ldots a_n}(\theta_1 \ldots \theta_n)\ =\ \langle vac|\mathcal O(0)|\theta_1 \ldots \theta_n\rangle_{a_1 \ldots a_n},
\end{equation}
where $\langle vac|$ is the vacuum state (without particles) of the scattering theory. 

Let us review general conditions providing a systematic way to reconstruction of the form factors. For the scalar operator $\mathcal O(x)$ the requirement for relativistic invariance implies  that its form factors depend on the rapidity differences $\theta_i-\theta_j$ only, while for the operator with the spin $s$ one gets the condition
\begin{equation}\label{ST:Frel}
	F^{\mathcal O}_{a_1 \ldots a_n}(\theta_1+\Lambda,\ldots,\theta_n+\Lambda)=e^{s\Lambda}F^{\mathcal O}_{a_1 \ldots a_n}(\theta_1,\ldots,\theta_n).
\end{equation}
The form factors~\eqref{ST:elFF} satisfy a set of requirements, called the form factor axioms, which are supplemented by the assumption of maximum analyticity~\cite{BKW}. They are the following,
\begin{enumerate}
	\item Watson's theorem,
		\begin{equation}\label{ST:mon1}
			F^{\mathcal O}_{\ldots a_k a_{k+1} \ldots}(\theta_1, \dots,\theta_k \theta_{k+1},\dots \theta_n)=S(\theta_{k+1}-\theta_k)F^{\mathcal O}_{\ldots a_{k+1} a_k \ldots}(\theta_1,\dots,\theta_{k+1},\theta_k,\dots \theta_n).
		\end{equation}
	\item The crossing-symmetry condition,
		\begin{equation}\label{ST:mon2}
			F^{\mathcal O}_{a_1 a_2 \ldots a_n}(\theta_1,\theta_2,\dots,\theta_n)=F^{\mathcal O}_{a_2 \ldots a_n a_1}(\theta_2,\dots,\theta_n,\theta_1+2\pi i).
		\end{equation}
	\item The kinematical pole condition,
			\begin{multline}\label{ST:kpole}
				-i\lim_{\theta'\to\theta}(\theta'-\theta)F^{\mathcal O}_{aaa_1\ldots a_n}(\theta'+i\pi,\theta,\theta_1,\dots,\theta_n)=\\
				=\Bigl(1-\prod_{j=1}^n S_{aa_j}(\theta-\theta_j)\Bigr)F^{\mathcal O}_{a_1\ldots a_n}(\theta_1,\ldots,\theta_n).
			\end{multline}
	\item The bound state pole condition,
			\begin{multline}\label{ST:bpole}
						-i\lim_{\theta'\to\theta}(\theta'-\theta)F^{\mathcal O}_{a_1 \ldots bc\ldots a_n}(\theta_1,\ldots,\theta'+i\bar u_{bd}^{c}, \theta-i\bar u_{cd}^{b},\ldots, \theta_n)=\\
=\Gamma_{bc}^d F^{\mathcal O}_{a_1 \ldots d\ldots a_n}(\theta_1,\dots,\theta,\ldots,\theta_n),
			\end{multline}
where $\bar u_{ab}^c=\pi-u_{ab}^c$.
\end{enumerate}

These axioms do not specify the dependence of the form factors on the operator $\mathcal O(x)$ and further requirements are necessary to identify the form factors of a specific operator among all solutions to the form factor axioms. This is in general a non-trivial problem. However there are the two useful criteria. The first one argued in~\cite{DM1} restricts the asymptotic behavior of the form factor. For the scaling operator $\mathcal O(x)$ of the scaling dimension $2\Delta_{\mathcal O}$ the form factor growth for large values of rapidities is bounded by
\begin{equation}\label{ST:ubound}
	F^{\mathcal O}_{a_1 \ldots a_n}(\theta_1,\dots,\theta_N)\leq e^{\Delta_{\mathcal O}|\theta_i|},\qquad {\rm as }\ |\theta_i|\rightarrow\infty.
\end{equation}
The second criterion essentially depends on the normalization of the form factors. For later use let us choose the following normalization,
\begin{equation}\label{ST:fdef}
	F^{\mathcal O}_{a_1 \ldots a_n}(\theta_1,\dots,\theta_N)=\langle \mathcal O\rangle f^{\mathcal O}_{a_1 \ldots a_n}(\theta_1,\dots,\theta_N),
\end{equation}
where $\langle \mathcal O\rangle$ is the vacuum expectation value of the local operator $\mathcal O$. This choice of the normalization is consistent with the spectral representation of the two-point correlation function
\begin{equation}
	\langle\mathcal O(x)\mathcal O(0)\rangle=\sum_{n=0}^{\infty}\sum_{\{a_n\}}\frac{1}{n!}\frac{d\theta_{1}}{2\pi}\ldots \frac{d\theta_n}{2\pi} |F^{\mathcal O}_{a_1 \ldots a_n}(\theta_1,\dots,\theta_n)|^2e^{-x\sum m_j\cosh\theta_j}
\end{equation}
Further we refer to the function $f^{\mathcal O}_{a_1,\ldots,a_n}(\theta_1,\dots,\theta_N)$ defined in~\eqref{ST:fdef} as to the non-normalized form factor. In these notation the second criterion proposed in~\cite{DSC} can be formulated in the following way. The form factors of the spineless exponential operators possess the factorization property, namely,
\begin{equation}\label{ST:factor}
	f^{\mathcal O}_{a_1\ldots a_n}(\theta_1+\Lambda,\ldots,\theta_m+\Lambda,\theta_{m+1},\dots \theta_N)=f^{\mathcal O}_{a_1\ldots a_m}(\theta_1,\dots, \theta_m)f^{\mathcal O}_{a_{m+1} \ldots a_n}(\theta_{m+1} \dots \theta_n),
\end{equation}
as $\Lambda\to\infty$ for any $m\in(0,\dots,n)$. This property had already been noticed to be satisfied in the solutions of many models and is believed to be a distinguishing property of exponential operators.

The solution to the system of the equations~\eqref{ST:mon1}-\eqref{ST:bpole} can be represented in the convenient form by using minimal two-particle form factors. The minimal two-particle form factor, denoted by $R_{ab}(\theta)$, is a solution to the equations~\eqref{ST:mon1} and~\eqref{ST:mon2} being specified to the case $n=2$,
\begin{equation}
	R_{ab}(\theta)=S_{ab}(\theta)R_{ba}(-\theta),\qquad R_{ab}(i\pi+\theta)=R_{ab}(i\pi-\theta).
\end{equation}
In order to determine an unique solution to these equations one imposes further restrictions, namely, $R_{ab}(\theta)$ is an analytic function in the ``physical strip'' and has nor zeros nor poles in this range. By using these notation the $n$-particle solution to the form factor axioms can be represented in the following form,
\begin{equation}
	f^{\mathcal O}_{a_1\ldots a_n}(\theta_1,\ldots,\theta_n)=J^{\mathcal O}_{a_1\ldots a_n}(e^{\theta_1},\ldots,e^{\theta_n})\prod_{1\leq i<j\leq n}R_{a_ia_j}(\theta_i-\theta_j).
\end{equation}
Here the function $J^{\mathcal O}_{a_1\ldots a_n}(e^{\theta_1},\ldots,e^{\theta_n})$ is a symmetric and $2\pi i$-periodic function in the variables $\theta_i$ containing all expected kinematical and bound state poles. This function represents a solution to the second pair of the form factor axioms~\eqref{ST:kpole} and~\eqref{ST:bpole}. Further we refer to these equation as to the bootstrap ones. The whole set of the form factors can be uniquely determined by the form factors which involve ``fundamental particles'' only. We referred to the particle $A_1$ as to the fundamental one if all other particles of the model $A_{a}$, $a=1,\ldots, n$ can be obtained as the ``bound states'' of some number of $A_1$. As the form factors involving ``fundamental particles'' are calculated the other form factors can be obtained by using the residue equation~\eqref{ST:bpole}. Therefore we restrict our attention to the calculation of the multi-particle form factors.

Further we consider form factors of the exponential operators only. In this case in framework of the free-field representation one can reduce their calculation to a combinatoric procedure~\cite{Lukyanov}. In the next section we briefly review the free-field representation for the form factors of the exponential operators in the Bullough-Dodd model.

\section{The Bullough-Dodd model}\label{BD}
In this section we briefly describe a scattering theory of the Bullough-Dodd model~\cite{BD}. This model is a two-dimensional integrable quantum field theory defined by the Euclidean action
\begin{equation}\label{BD:S}
	S_{BD}=\int d^2x\biggl(\frac{1}{16\pi}(\partial_\nu\varphi)^2+\mu(e^{\sqrt{2}b\varphi}+2e^{-\frac{b}{\sqrt{2}}\varphi})\biggr),
\end{equation}
where $b$ is the coupling constant and $\mu$ is the regularized mass parameter which is related to the mass $m$ of the single bosonic particle $A$ in the spectrum of the model in the following,
\begin{equation}
	m=\mu^\frac{1}{2Qb}\frac{2\sqrt3 \Gamma\bigl(\frac13\bigr)}{\Gamma\bigl (1+\frac{b}{3Q}\bigr) \Gamma\bigl( \frac{1}{3Qb}\bigr)}\Bigl(-\frac{\pi \Gamma(1+2b^2)}{\Gamma(-2b^2)} \Bigr)^\frac{1}{6Qb}\Bigl( -\frac{2\pi \Gamma(1+b^2/2)}{\Gamma(-b^2/2)} \Bigr)^\frac{1}{3Qb}.
\end{equation}
Here we introduce the convenient notation
\begin{equation}
Q=b^{-1}+b.
\end{equation}
The Bullough-Dodd model possesses an infinite number of local conserved charges of the odd integer spin $s$ except multipliers of $3$,
\begin{equation}\label{BD:spin}
	s=1,5,7,11,13,\ldots.
\end{equation}
The integrability of the model implies that the $n$-particle $S$-matrix factorizes into the $n(n-1)/2$ two-particle scattering amplitudes,
\begin{equation}\label{BD:Smatrix}
	S(\theta)=\frac{\tanh\frac{1}{2}(\theta+\frac{2i\pi}{3})\tanh \frac{1}{2}(\theta-\frac{2i\pi}{3bQ}) \tanh\frac{1}{2}(\theta-\frac{2i\pi b}{3Q})}{\tanh\frac{1}{2}(\theta-\frac{2i\pi}{3}) \tanh\frac{1}{2}(\theta+\frac{2i\pi}{3Qb})\tanh\frac{1}{2}(\theta+\frac{2i\pi b}{3Q})}.
\end{equation}
For real values of the coupling constant $b$ the $S$-matrix has a simple pole at $\theta=2i\pi/3$ corresponding to the bound state represented by the particle $A$ itself in the scattering processes
\begin{equation}\label{BD:Scaterring}
	A\times A \to A \to A\times A.
\end{equation}
The space of local operators of the model consist of the exponential operators
\begin{equation}\label{BD:V}
	V_a(x)=e^{a\varphi(x)}
\end{equation}
and their descendants. We are interested in the form factors of the exponential operators only. Let us label them in the following,
\begin{equation}\label{BD:Fa}
	F^a(\theta_1,\ldots,\theta_N)=\langle vac|V_a(x)|\theta_1,\ldots,\theta_N\rangle.
\end{equation}
By using free-field representation one can obtain a set of functions which satisfy the form factor axioms together with the requirements~\eqref{ST:ubound} and~\eqref{ST:factor}. Consequently these functions can be identified with the form factors of the exponential operators. We briefly review the free-field construction in the Bullough-Dodd model~\cite{BL}. We do not need the detailed description in terms of the auxiliary free-field. Thus, let us formulate the result only.

Consider a pair of the operators $\Lambda(\theta)$ and $\Lambda^{-1}(\theta)$ and define the braces $\langle\langle\ldots\rangle\rangle$ by
\begin{equation}\label{BD:LL}
	\begin{aligned}
	\langle\langle\Lambda(\theta)\rangle\rangle&=1,\\
	\langle\langle \Lambda^{\sigma'}(\theta')\Lambda^\sigma(\theta)\rangle\rangle&=\left [R(\theta-\theta')\right]^{\sigma'\sigma},\quad \sigma',\sigma=\pm1,
	\end{aligned}
\end{equation}
where $R(\theta$) is the two-point minimal form factor of the Bullough-Dodd model~\cite{FMS,A}. It is convenient to use the following parametrization for this function,
\begin{equation}\label{BD:R}
	R(\theta)=\mathcal N(b)\frac{g_0(\theta)g_\frac{2}{3}(\theta)}{g_\frac{2b}{3Q}(\theta)g_\frac{2}{3Qb}(\theta)}.
\end{equation}
Here we introduce the functions $g_\alpha(\theta)$ defined by
\begin{equation}\label{IMMF:g}
	g_\alpha(\theta)=\exp\Bigl(2\int_0^\infty\frac{dt}{t}\frac{\cosh\frac{(2\alpha-1)t}{2}}{\cosh\frac{t}{2}\sinh t}\sin^2\frac{(i\pi-\theta)t}{2\pi}\Bigr).
\end{equation}
The normalization constant $\mathcal N(b)$ is chosen such that $R(\theta)\to1$ as $\theta\to\infty$, namely,
\begin{equation}
	\mathcal N(b)=\exp\Bigl(-4\int_0^\infty\frac{dt}{t}\frac{\cosh\frac{t}{6}\sinh\frac{tb}{3Q}\sinh\frac{t}{3Qb}}{\sinh t\cosh\frac{t}{2}}\Bigr).
\end{equation}
By construction the braces $\langle\langle\ldots\rangle\rangle$ with arbitrary number of the operators $\Lambda(\theta)$ and $\Lambda^{-1}(\theta)$ factorizes into the pair braces~\eqref{BD:LL},
\begin{equation}\label{BD:LLdef}
	\langle\langle\Lambda^{\sigma_N}(\theta_N)\ldots\Lambda^{\sigma_1}(\theta_1)\rangle\rangle=\prod_{1\leq i<j\leq N}\langle\langle \Lambda^{\sigma_i}(\theta_i)\Lambda^{\sigma_j}(\theta_j)\rangle\rangle.
\end{equation}
With these notation Lukyanov's generators are given by
\begin{equation}\label{BD:T}
	T(\theta)=\rho\ \Bigl(\gamma\Lambda(\theta+\frac{i\pi}{2})+\gamma^{-1}\Lambda^{-1}(\theta-\frac{i\pi}{2})+ h :\Lambda(\theta+\frac{i\pi}{6})\Lambda(\theta-\frac{i\pi}{6}):\Bigr),
\end{equation}
where the $\theta$-independent constants $\gamma$, $h$, $ \rho $ are the following,
\begin{equation}\label{BD:const}
	\begin{aligned}
	&\gamma=-i\exp\Bigl(\frac{i\pi}{6Q}(4\sqrt{2}a-b+b^{-1})\Bigr),\\
	& h =2\sin\frac{\pi(b-b^{-1})}{6Q},\\
	&\rho=\sqrt{\frac{\sin\frac{\pi}{3}}{\sin\frac{2\pi b}{3Q}\sin\frac{2\pi}{3Qb}}}\exp\Bigl(2\int_0^\infty\frac{dt}{t}\frac{\cosh\frac{t}{6}\sinh\frac{tb}{3Q}\sinh\frac{t}{3Qb}}{ \sinh t\cosh\frac{t}{2}}\Bigr).
	\end{aligned}
\end{equation}
Notice that these notation differs from those one of~\cite{BL} by the substitution $b\to b^{-1}$. The free-field construction possesses the duality transformation $b\leftrightarrow b^{-1}$ which is also a symmetry of the $S$-matrix. However, this substitution is applied to adopt the conventional notation of the conformal field theory. As a result, the form factors of the exponential operators can be represented as follows,
\begin{equation}\label{BD:F}
	F^a(\theta_1,\ldots,\theta_N)\equiv\langle e^{a\varphi}\rangle f^a(\theta_1,\ldots,\theta_N)=\langle e^{a\varphi}\rangle\langle\langle T(\theta_N) \ldots T(\theta_1)\rangle\rangle,
\end{equation}
where $\theta$-independent factor $\langle  e^{a\varphi}\rangle$ is the vacuum expectation value of the exponential operator in the Bullough-Dodd model~\cite{FLZZ}. By using~\eqref{BD:LL} one can show that the non-normalized form factors $f^a(\theta_1,\ldots,\theta_N)$ in~\eqref{BD:F} possess the following parametrization,
\begin{equation}\label{BD:Jdef}
	f^{a}(\theta_1,\ldots,\theta_n)=\rho^NJ_{N,a}(x_1,\ldots,x_N) \prod_{1\leq i<j\leq N} R(\theta_i-\theta_j).
\end{equation}
Here $J_{N,a}(x_1,\ldots,x_N)$ is a symmetric rational function in the variables $x_i=e^{\theta_i}$ with proper kinematical and dynamical poles. Notice that in the proposed free-field representation the calculation of the form factors reduces to a  combinatoric procedure. In the next section we obtain the free-field representation for the functions $J_{N,a}(x_1,\ldots,x_N)$ alone. As a consequence we propose particular recurrence relations between them. 

\subsection{Free-field representation for the functions $J_{N,a}(x_1,\ldots,x_n)$}
We follow the guideline of Lukyanov's construction of the free-field representation~\cite{BL}. Let us specify the differences. We consider the Heisenberg algebra generated by a countable set of the elements $a^s_n$, $n\in \mathbb Z,\ n\neq0$ and $s=\pm,0$, with the following commutation relations,
\begin{equation}\label{BD:acomm}
	[a_n^{s},a_m^{s'}]=-4n\sin\Bigl(\frac{(n-1)\pi b}{3Q}+\frac{\pi}{6}-\frac{n\pi}{3}\Bigr)\sin\Bigl(\frac{(n-1)\pi b}{3Q}+\frac{\pi}{6}\Bigr)A^{s,s'}_n\delta_{n,-m},
\end{equation}
and the matrix $A^{\sigma,\sigma'}$ is given by
\begin{equation}
A_{n}^{s,s'}=	\begin{pmatrix*}[l]
				0	&	\omega^n(1+\omega^n)	&	\omega^n\\
				\omega^{-n}(1+\omega^{-n})	&	0	&\omega^{-n}\\
				\omega^{-n}	&	\omega^n 	&	1
				\end{pmatrix*},\qquad \omega=\exp\Bigl(\frac{i\pi}{3}\Bigr),
\end{equation}
where the raw $s=+,-,0$ and the column $s'=+,-,0$. Let us define the exponential operators
\begin{equation}\label{BD:lambda}
	\lambda^s(z)=\ :\exp\sum\limits_{i\neq0}\frac{a^s_n}{n}z^{n}:\ ,\quad s=\pm,0.
\end{equation}
Here the normal ordering is not needed for the operators $\lambda^\pm(z)$ since the corresponding generators commute. Following~\cite{BL} we can determine the proper Lukyanov's generators by using the operators~\eqref{BD:lambda} and identify the exponential fields with the projectors on the corresponding Fock modules. It is convenient to formulate the result as a following prescription. The function $J_{N,a}(x_1,\ldots,x_N)$ can be represented as a matrix element,
\begin{equation}\label{BD:J}
	J_{N,a}(x_1,\ldots,x_N)=\langle t(x_1)\dots t(x_N) \rangle,
\end{equation}
where
\begin{equation}\label{BD:t}
t(z)=\gamma\lambda^+(z)+\gamma^{-1}\lambda^-(z)+ h \lambda^0(z)\ .
\end{equation}
Here $\gamma$ and $ h $ are defined in~\eqref{BD:const}. In~\eqref{BD:J} we use the notation
\begin{equation}
\langle\ldots\rangle=\langle 0|\ldots|0\rangle
\end{equation}
for the corresponding matrix elements and the vacuums are defined by the relations
\begin{equation}
	a^s_n|0\rangle=0,\quad \langle 0|a^s_{-n}=0,\quad s=\pm,0,\quad\text{for}\ n>0.
\end{equation}
The calculation of the matrix elements multi-particle matrix elements is performed using Wick's averaging procedure with the following rules,
\begin{equation}\label{BD:ll}
	\begin{aligned}
	\lambda^{\pm}(x)\lambda^{\pm}(x')\ =&\ :\lambda^{\pm}(x)\lambda^{\pm}(x'):\ ,\\
	\lambda^0(x)\lambda^0(x')\ =&\ f\Bigl(\frac{x}{x'}\Bigr) :\lambda^0(x)\lambda^0(x'):\ ,\\
	\lambda^0(x')\lambda^{\pm}(x)\ =\ \lambda^{\pm}(x)\lambda^0(x')\ =&\ f\Bigl(\frac{x}{x'}\omega^{\pm1}\Bigr):\lambda^{\pm}(x)\lambda^0(x'):\ ,\\
	\lambda^{-}(x')\lambda^{+}(x)\ =\ \lambda^{+}(x)\lambda^{-}(x')\ =&\ f\Bigl( \frac{x}{x'}\omega\Bigr)f\Bigl( \frac{x}{x'}\omega^2 \Bigr):\lambda^{+}(x)\lambda^{-}(x') :\ ,
	\end{aligned}
\end{equation}
where the function $f(x)$ is given by
\begin{equation}\label{BD:f}
	f(x)=\frac{x+x^{-1}+h^2-2}{x+x^{-1}-1}.
\end{equation}
The free-field representation~\eqref{BD:J} is very similar to Lukyanov's representation~\eqref{BD:F}. Its main advantage is that redundant factors of the two-particles minimal form factors are excluded from the construction. Therefore, calculation of the matrix elements involve Wick's averaging represented by rational functions in the variables $x_i$. Notice that the averaging procedure~\eqref{BD:ll} in the $N$-particle matrix elements generates $3^N$ terms. The particular terms may possess poles at the points $x_i=x_j\omega^{\pm2}$ corresponding to bound state poles, at $x_i=-x_j$ corresponding to kinematical poles, and redundant ones at $x_i=x_j\omega^{\pm1}$. The presence of the redundant poles complicates calculation of the form factors. However, in Appendix~\ref{BDA} we prove that their contributions in the whole expression for the matrix elements vanish.

The simple analytic structure of the matrix elements allows us to obtain recurrence relations between them. Consider the function $J_{N+1,a}(z,x_1,\ldots,x_N)$ as an analytic function in the variable $z$ depending on the parameters $x_1,\ldots,x_N$. One can separate the contribution of the poles from the regular part. Since the residues of this function can be evaluated explicitly (see Appendix~\ref{BDA}) the only unknown is the regular part. However, for the exponential operators the regular part of this function can be determined using the factorization property~\eqref{ST:factor}. We present details of the construction in Appendix~\ref{BDB}. As a result we can formulate the following proposition,
\begin{proposition}\label{BD:prop}
The recurrence relations
\begin{multline}\label{BD:ThR}
	\begin{aligned}
		J_{N+1,a}(z,x_1,\ldots,x_N)\ =\ &J_{1,a}J_{N,a}(x_1,\ldots,x_N)+\\
		&+\sum\limits_{n=1}^N\frac{x_n}{z+x_n}K_n(x_1,\ldots,x_N)J_{N-1,a}(x_1,\ldots,\cancel{x}_n,\ldots,x_N)+\\
		&+\sum\limits_{n=1}^N\frac{x_n\omega^{2}}{z-x_n\omega^2}B_n^+(x_1,\ldots,x_N)J_{N,a}(x_1,\ldots,x_n\omega,\ldots,x_N)-\\
		&-\sum_{n=1}^N\frac{x_n\omega^{-2}}{z-x_n\omega^{-2}}B_n^-(x_1,\ldots,x_N)J_{N,a}(x_1,\ldots,x_n\omega^{-1},\ldots,x_N).
	\end{aligned}
\end{multline}
together with the initial condition
\begin{equation}\label{BD:J1}
	J_{1,a}\equiv J=4\sin\Bigl(\frac{\pi\sqrt2a}{3Q}\Bigl)\cos\Bigl( \frac{\pi}{6Q}(2\sqrt2a -b+b^{-1})\Bigr),
\end{equation}
uniquely define the set of functions $J_{N,a}(x_1,\ldots,x_N)$ of exponential operators. The functions $K_n(x_1,\ldots,x_N)$ and $B_n^\pm(x_1,\ldots,x_N)$ are given by
\begin{equation}\label{BD:BK}
	\begin{aligned}
		K_n(x_1,\ldots,x_N)=\ &\frac{(h^2-3)(h^2-1)}{2(\omega-\omega^{-1})}\biggl(\prod_{i\neq n}f\Bigl(\frac{x_n}{x_i}\omega^{2}\Bigr)f\Bigl(\frac{x_n}{x_i}\omega\Bigr)-\prod_{i\neq n}f\Bigl(\frac{x_n}{x_i}\omega^{-2}\Bigr)f\Bigl(\frac{x_n}{x_i}\omega^{-1}\Bigr)\biggr),\\
		B_n^{\pm}(x_1,\ldots, x_N)=\ &\frac{ h ( h ^2-1)}{\omega-\omega^{-1}}\prod_{i\neq n} f\Bigl( \frac{x_n}{x_i}\omega^{\pm1} \Bigr).
	\end{aligned}
\end{equation}
\end{proposition}
As an example of possible applications of the recurrence relations~\eqref{BD:ThR} in Appendix~\ref{EM} we prove that the form factors of the exponential operators satisfy the quantum equation of motion. Another example is provided by the so called reflection relations~\cite{FLZZ}.

\subsection{Reflection relations for exponential operators}
The Bullough-Dodd model can be interpreted as the perturbed Liouville theory in the two different ways, with either the first exponent in~\eqref{BD:S} or the second one is taken as the perturbing operator. Let us introduce the notation
\begin{equation}\label{BD:Q}
	Q_L=\frac{1}{\sqrt2b}+\sqrt2b,\quad Q'_L=\frac{\sqrt2}{b}+\frac{b}{\sqrt2}
\end{equation}
There is a conjecture based on the features of the Liouville theory that the form factors of the exponential operators in the Bullough-Dodd model satisfy the relations~\cite{FLZZ},
\begin{equation}\label{BD:Cref}
	\begin{gathered}
	\langle vac|e^{a\varphi}|\theta_1,\ldots\theta_N\rangle=R(a)\langle vac| e^{(Q_L-a)\varphi}|\theta_1,\ldots,\theta_N\rangle,\\
	\langle vac|e^{-a\varphi}|\theta_1,\ldots,\theta_N\rangle=R'(a)\langle vac|e^{(-Q'_L+a)\varphi}|\theta_1,\ldots,\theta_N\rangle,
	\end{gathered}
\end{equation}
where the explicit expression for the reflection amplitude $R(a)$ is presented in~\cite{FLZZ} and $R'(a)$ is obtained from that by the substitutions $\mu\to2\mu$ $b\to b/2$. The vacuum expectation values possess the same relations. Consequently, for the non-normalized form factors defined in~\eqref{ST:fdef} the reflection relations can be presented in the following form,
\begin{equation}\label{BD:fref}
	\begin{gathered}
	f^{a}(\theta_1,\ldots,\theta_N)=f^{Q_L-a}(\theta_1,\ldots,\theta_N),\\
	f^{-a}(\theta_1,\ldots,\theta_N)= f^{-Q'_L+a}(\theta_1,\ldots,\theta_N).
	\end{gathered}
\end{equation}
By construction~\eqref{BD:Jdef} the non-normalized form factors only depend on the parameter $a$ through the functions $J_{N,a}(x_1,\ldots,x_N)$ whose dependence on this parameter is completely determined by the initial conditions to the recurrence relations~\eqref{BD:J1}. One can easily check that the functions $J_{1,a}$ indeed satisfy the reflection relations. This proves~\eqref{BD:fref}.

\subsection{An explicit calculation of the form factors}
From the representation~\eqref{BD:J} it follows that the functions $J_{N,a}(x_1,\ldots,x_N)$ are symmetric and $2\pi i$-periodic functions in the variables $x_i=e^{\theta_i}$. Consequently, these functions can be expressed in the convenient form by means of symmetric functions. A basis in the space of the symmetric functions in $N$ variables is provided by the elementary symmetric polynomials $\sigma_N(x_1,\ldots,x_N)$ defined by the generating function
\begin{equation}
	\prod_{i=1}^N(x+x_i)=\sum_{i=0}^Nx^{N-i}\sigma_i(x_1,\ldots,x_N).
\end{equation}
It is convenient to separate the poles contribution prescribed by the recurrence relation~\eqref{BD:ThR} into an overall factor. Then the functions $J_{N,a}(x_1,\ldots,x_N)$ can be parametrized as
\begin{equation}\label{BD:Jparam}
	J_{N,a}(x_1,\ldots,x_N)=\Lambda_N(\sigma_1,\ldots,\sigma_N)\prod_{1\leq i<j\leq N}\frac{1}{(x_i+x_j)(x_i^2+x_ix_j+x_j^2)},
\end{equation}
where $\Lambda_N(\sigma_1,\ldots,\sigma_N)$ are symmetric polynomials in the variables $\sigma_i(x_1,\ldots,x_N)$. Taking into account~\eqref{ST:Frel} one can show that for spinless operators the total degree of this polynomial is equal to $3N(N-1)/2$ while the partial degree in the each variable $\sigma_i$ is bounded by the the condition~\eqref{ST:ubound} and could not exceed $3(N-1)$. In the framework of the free-field representation~\eqref{BD:J} we compute the polynomials $\Lambda_N(\sigma_1,\ldots,\sigma_N)$ up to $N=4$,
\begin{equation}
	\begin{aligned}
	\Lambda_1=&\ J,\\
	\Lambda_2(\sigma_1,\sigma_2)=&\ J^2 \sigma_1^2-J(J+h-h^3 )\sigma_2,\\
	\Lambda_3(\sigma_1,\sigma_2,\sigma_3)=&\ \ \ \ J(J^2+ (-3+ h  ^2 )  (-1+ h  ^2 )^2+J  ( h  - h  ^3 ) )\sigma_2^3\sigma_3\\
	&+  J(J^2+ (-3+ h  ^2 )  (-1+ h  ^2 )^2+J  ( h  - h  ^3 ) )\sigma_1\sigma_3^2\\
	&- J h    (J+ h  - h  ^3 )  (4-5  h  ^2+ h  ^4 )\sigma_1\sigma_2\sigma_3^2\\
	&-J^2  (J+ h  - h  ^3 )\sigma_1\sigma_2\\
	&-J^2  (J+ h  - h  ^3 )\sigma_1^4\sigma_2\sigma_3\\
	&+ J (-J^2+3  (-1+ h  ^2 )^2+J  h    (4-5  h  ^2+ h  ^4 ) )\sigma_1^2\sigma_2^2\sigma_3\\
	&+J^3\sigma_1^3\sigma_2^3,
	\end{aligned}
\end{equation} 
and explicit expression for the four-particle symmetric polynomial $\Lambda_4(\sigma_1,\sigma_2,\sigma_3,\sigma_4)$ is presented in Appendix~\ref{4p}.

As a simple check of the proposed expressions let us consider form factors of the exponential operators which appear in the action~\eqref{BD:S}. The general expression of the stress-energy tensors trace $\Theta(x)$ which is compatible with the quantum equation of motion is given by the linear combination of these operators~\cite{MS}. Taking into account the stress-energy tensor conservation law one can show that the $N$-particle form factors of $\Theta(x)$ has to be proportional to the combination of symmetric polynomials $\sigma_1\sigma_{N-1}$ for $N>2$. Using explicit expressions of the three- and the four-particle form factors we checked  that the form factors of the corresponding exponential operators indeed possess this property.

\section{The $\Phi_{1,2}$ perturbed minimal models}\label{cBD}
Consider an imaginary coupling affine Toda theory based
on the twisted affine Kac-Moody algebra $A_{2}^{(2)}$, which is obtained from~\eqref{BD:S} by the substitutions $b\to i\beta$, $\mu\to-\mu$, 
\begin{equation}\label{cBD:S}
	S_{cBD}=\int\biggl(\frac{1}{16\pi}(\partial\varphi)^2-\mu(e^{i\sqrt{2}\beta\varphi}-2e^{-i\frac{\beta}{\sqrt{2}}\varphi})\biggr).
\end{equation}
This model can be referred to as the Zhiber-Mikhailov-Shabat or the imaginary coupling Bullough-Dodd model. It is not unitary theory since its Hamiltonian is not hermitian. However in~\cite{Smirnov} it was shown that the model is reducible for certain values of its coupling constant.
Let us briefly describe the restriction of the complex Bullough-Dodd model. By using the action~\eqref{cBD:S} it is possible to determine non-local conserved charges that commute with the Hamiltonian. These charges generate the quantum affine algebra $U_q(A_{2}^{(2)})$ with
\begin{equation}\label{cBD:q}
	q=e^\frac{i\pi}{2\beta^2}
\end{equation}
and fix the $S$-matrix of the model up to a scalar factor~\cite{BerLe,Smirnov,Eft}. The restriction proceeds in the following way. The quantum algebra $U_q(A_{2}^{(2)})$ contains two subalgebras $U_q(sl_2)$ and $U_{q^4}(sl_2)$ and both of them can be used to obtain quantum group restrictions of~\eqref{cBD:S}. Consider the first possibility. Let the parameter $2\beta^2$ takes the following rational values
\begin{equation}\label{cBD:beta}
2\beta^2=\frac{p}{p'}
\end{equation}
where $p$ and $p'$ are relatively primary integers such that $p'>p>1$. In this case the Hilbert space of the imaginary coupling Bullough-Dodd model can be consistently truncated to the representations of $U_q(sl_2)$. The details of the construction can be found in~\cite{Smirnov,Eft}. The restricted theory is known to coincide with the $\Phi_{1,2}$ perturbations of minimal models $\mathcal M_{p,p'}$~\cite{Smirnov} whose formal action can be represented in the following form,
\begin{equation}\label{cBD:MM}
	S_{\mathcal M_{p,p'}}+\lambda\int d^2x\Phi_{1,2}(x).
\end{equation}
The relation between the corresponding coupling constants $\mu$ and $\lambda$ can be established using the results of~\cite{FateevB324},
\begin{equation}\label{cBD:lambda}
	\lambda^2=-\frac{4\pi \mu^3}{(4\beta^2-1)^2}\frac{\Gamma^2(1-2\beta^2)\Gamma(4\beta^2)}{\Gamma^2(2\beta^2)\Gamma(1-4\beta^2)}.
\end{equation}
Let us briefly introduce conventional notation used in the conformal field theory. The central charge of the minimal model $\mathcal M_{p,p'}$ is given by
\begin{equation}\label{cBD:MMc}
	c=1+6Q_L^2,
\end{equation}
where $Q_L$ was introduced in~\eqref{BD:Q} and the substitution $b=i\beta$ is assumed. The primary operator content of the minimal model $\mathcal M_{p,p'}$ is given by a set of the degenerate fields $\Phi_{m,n}$, with $m=1,\ldots,p-1$ and $n=1,\ldots,p'-1$, whose conformal dimensions can be parametrized in the following, 
\begin{equation}\label{cBD:Delta}
	\Delta_{m,n}=a_{m,n}(Q_L-a_{m,n}),
\end{equation}
where
\begin{equation}
a_{m,n}=-\frac{(n-1)}{2}\sqrt2i\beta-\frac{(m-1)}{2}\frac{1}{\sqrt2i\beta}.
\end{equation}
The normalization of the primary operators is fixed by the two-point function,
\begin{equation}\label{cBD:N}
	\langle \Phi_{m,n}(x)\Phi_{m,n}(0)\rangle=|x|^{-4\Delta_{m,n}}.
\end{equation}
The primary operators of the perturbed minimal models relates with the specific exponential fields of the imaginary coupling Bullough-Dodd model~\cite{RS,BerLe}. Namely, the following exponential operators,
\begin{equation}
	V_{a_{1,n}}(x)=e^{a_{1,n}\varphi(x)},
\end{equation}
commute with particular generators of the $U_q(sl_2)$ and can be identified with the primary operators $\Phi_{1,n}$ of the perturbed minimal models.

The fundamental particle of the complex Bullough-Dodd model is a kink triplet. The relation between masses of the lightest kinks and the coupling constant $\lambda$ is established in~\cite{FateevB324}.  Under the action $U_q(sl_2)$ the fundamental kinks form either singlet bound states generating a series of breathers or triplet ones corresponding to higher kinks. After analytical continuation to the imaginary values of the coupling constant and quantum group restriction the $S$-matrix of the Bullough-Dodd model can be treated  as the lightest particles $S$-matrix in the $\Phi_{1,2}$ perturbed minimal models~\cite{Koubek}. It develops two additional poles in the ``physical strip'' corresponding to heavier particles. We propose an identification between the form factors of the Bullough-Dodd model and the lightest particle form factors of the corresponding minimal models. One can check that after analytical continuation to the imaginary values of the coupling constant the form factors~\eqref{BD:Jdef} satisfy the form factor axioms. In particular, the two-point minimal form factor develops two additional poles in the ``physical strip'' whose positions correspond to the poles of the $S$-matrix of the lightest particles and can be explicitly determined by using the following functional relations,
\begin{equation}\label{cBD:g}
	\begin{gathered}
	g_{1+\alpha}(\theta)=g_{-\alpha}(\theta),\\
	g_\alpha(\theta)g_{-\alpha}(\theta)=\mathcal P_\alpha(\theta)\equiv\frac{\cos\pi\alpha-\cosh\theta}{2\cos^2\frac{\pi\alpha}{2}}.
	\end{gathered}
\end{equation}
Further, we restrict our attention to the non-normalized form factors only. In the free-field representation the correct normalization is fixed by the vacuum expectation value of the corresponding operator. However, perturbed minimal models possess several vacuums and calculation of their contribution to the vacuum expectation values is an open problem. As a result let us propose the following representation for the lightest particles form factors of the $\Phi_{1,n}$ primary operators in the $\Phi_{1,2}$ perturbed minimal models $\mathcal M_{p,p'}$,
\begin{equation}\label{cBD:f}
f^{a_{1,n}}_{1\ldots1}(\theta_1,\ldots,\theta_N)=f^{a_{1,n}}(\theta_1,\ldots,\theta_N)\Bigr|_{2b^2=-p/p'}.
\end{equation}
Notice that the form factor in the right hand side of these relation admits the free-field representation~\eqref{BD:J}. 
Being analytically continued to the imaginary values of the coupling constant the minimal two-particle form factors develop additional poles. Therefore, the poles emerge in Lukyanov's representation~\eqref{BD:F} due to Wick's averaging procedure~\eqref{BD:LL}. However, in the proposed free-field representation~\eqref{BD:J} these redundant factors are excluded from the construction. The averaging procedure generates functions with the same analytic structure as in the Bullough-Dodd model. In the next section we consider the Ising model in a magnetic field as an significant example of the proposed free field representation.


\subsection{Ising model in a magnetic field}\label{IMMF}
The model~\eqref{cBD:MM} with $(p,p')=(3,4)$ describes the Ising model at critical temperature in non-zero magnetic field (IMMF). The minimal model $\mathcal M_{3,4}$ is characterized by its central charge $c=1/2$ and contains two non-trivial primary operators $\Phi_{1,2}$ and $\Phi_{1,3}$. Their conformal dimensions are given by~\eqref{cBD:N}, namely, $\Delta_{1,2}=1/16$ and $\Delta_{1,3}=1/2$. The Ising model can be defined in a different way, namely, as a scaling limit of the Ising lattice model at its critical point. Therefore, one can define the spin density operator $\sigma(x)$ and the energy density operator $\epsilon(x)$ as the scaling limit of the corresponding lattice operators. Assuming the normalization~\eqref{cBD:N} for these operators one can identify them with primary operators of the minimal model, namely, $\sigma(x)=\Phi_{1,2}(x)$ and $\epsilon(x)=\Phi_{1,3}(x)$. The coupling constant $\lambda$ defined in~\eqref{cBD:lambda} can be referred to as external magnetic field or, more precisely, as its scaling limit.

The spectrum of the model consist of eight different species of self-conjugated particles $A_i$, $i=1,\ldots,8$. Under analytical continuation to imaginary value of the coupling constant, $2b^2=-3/4$, the $S$-matrix of the Bullough-Dodd becomes the $S$-matrix of the lightest particles in the IMMF,
\begin{equation}\label{IMMF:S11}
	S_{11}(\theta)=\frac{\tanh\frac{1}{2}(\theta+\frac{2i\pi}{3})\tanh\frac{1}{2}(\theta+\frac{2i\pi}{5})\tanh\frac{1}{2}(\theta+\frac{i\pi}{15})}{\tanh\frac{1}{2}(\theta-\frac{2i\pi}{3})\tanh\frac{1}{2}(\theta-\frac{2i\pi}{5})\tanh\frac{1}{2}(\theta-\frac{i\pi}{15})}.
\end{equation}
The additional poles that emerges in the ``physical strip'' after analytical continuation of the $S$-matrix correspond to heavier particles. We denote the particle corresponding to the pole at $\theta=2\pi i/3$ by $A_2$ while those one corresponding to the pole at $\theta=i\pi/15$ by $A_3$. As shown in~\cite{ABZIMMF2} the full set of the two-particle amplitudes $S_{ab}(\theta)$, where $a,b=1,\ldots8$, can be reconstructed starting from the two-particle amplitude~\eqref{IMMF:S11}. Besides the bootstrap structure of the model fixes masses of all particles,
\begin{equation}\label{IMMF:m}
	\begin{aligned}
	& m_1=m,\quad &&m_2=2m\cos\frac{\pi}{5},\quad &&m_3=2m\cos\frac{\pi}{30},\\
	& m_4=2m_2\cos\frac{7\pi}{30},\quad &&m_5=2m_2\cos\frac{2\pi}{15},\quad && m_6=2m_2\cos\frac{\pi}{30},\\
	&m_7=4m_2\cos\frac{\pi}{5}\cos\frac{7\pi}{30},\quad &&m_8=4m_2\cos\frac{\pi}{5}\cos\frac{2\pi}{15},
	\end{aligned}
\end{equation}
Notice that under analytical continuation of the Bullough-Dodd model and quantum group restriction not all integral of motion survive. Instead of~\eqref{BD:spin} the IMMF possesses integral of motion with
\begin{equation}\label{IMMF:spin}
	s=1,7,11,13,17,19,23,29,\quad\text{mod 30}.
\end{equation}
The integers~\eqref{IMMF:spin} are are exactly the exponents of the Lie algebra $E_8$, repeated modulo $30$. In addition the number of particles is exactly the rank of $E_8$.

\subsection{Form factors in the IMMF model}
The form factors of the IMMF can be calculated in the framework of the proposed free-field representation. It is instructive to consider the analytic continuation of the form factors~\eqref{BD:Jdef} in detail. The analitical continuation of the two-particle minimal form factor can be performed by using relations~\eqref{cBD:g},
\begin{equation}\label{IMMF:Ran}
R(\theta)\Bigr|_{2b^2=- 3/4}=\frac{R_{11}(\theta)}{\mathcal P_{2/5}(\theta)\mathcal P_{1/15}(\theta)}
\end{equation}
where we introduce the notation
\begin{equation}\label{IMMF:R11}
R_{11}(\theta)=\mathcal N\biggl(i\sqrt\frac38\biggr)\ g_0(\theta)g_{2/3}(\theta)g_{2/5}(\theta)g_{1/15}(\theta).
\end{equation}
One can show that this function satisfy the form factor axioms~\eqref{ST:mon1},~\eqref{ST:mon2} and has nor zeros nor poles in the ``physical strip''. Therefore we refer to this function as to the minimal two-particle form factor of the lightest particles $A_1$ in the IMMF. The poles representing by the functions $\mathcal P_{\alpha}(\theta)$ in~\eqref{IMMF:Ran} correspond to the $A_2$ and $A_3$ bound state poles of the two-particle amplitude~\eqref{IMMF:S11}.

The analytic continuation of the function $J_{N,a}(x_1,\ldots,x_N)$ is straightforward. By construction this function satisfy the form factors axioms~\eqref{ST:bpole} and~\eqref{ST:kpole} at the dynamical poles, $\theta=2\pi i/3$, and kinematical ones respectively, has a correct large rapidity behavior~\eqref{ST:ubound} and possesses the factorization property~\eqref{ST:factor}. In the parametrization~\eqref{BD:Jdef} the lightest particle form factors in the IMMF can be represented in the following form,
\begin{equation}\label{IMMF:f}
	f^{a_{1,n}}_{1\ldots1}(\theta_1,\ldots,\theta_N)= \rho^N J_{N,a_{1,n}}(x_1,\ldots,x_N)\prod_{1\leq i<j\leq N}\frac{R_{11}(\theta_{ij})}{\mathcal P_{2/5}(\theta_{ij})\mathcal P_{1/15}(\theta_{ij})},
\end{equation}
where the substitutions $2b^2=-3/4$ is assumed. Notice that the functions $J_{N,a_{1,n}}(x_1,\ldots,x_N)$ in~\eqref{IMMF:f} are given by the corresponding matrix elements~\eqref{BD:J}. Therefore, one can obtain multi-particle form factors of the lightest particles either by using Wick's averaging procedure or by means of the proposed recurrence relations. We checked that both these methods up to four-particle form factors reproduces the results of~\cite{DelfinoM}, where the form factors were obtained by solving the bootstrap equations. Due to bootstrap structure of the model the form factors involving different species of particles can be obtained from~\eqref{IMMF:f} by means of the residue equation~\eqref{ST:bpole}. For example, in the proposed representation the computation of the four-particle form factor $f^{a_{1,n}}_{1,1,1,2}(x_1,x_2,x_3,x_4)$ reduces to the averaging procedure of the corresponding matrix element $J_{5,a_{1,n}}(x_1,x_2,x_4,x_5\omega^{3/5},x_5\omega^{-3/5})$. Notice that in this model the correct normalization of the form factors of primary operators can be obtained by using the vacuum expectation values of the corresponding exponential operators in the Bullough-Dodd model~\cite{FLZZ},
\begin{equation}
	\langle \Phi_{1,2}\rangle=-1.27758\ldots\ \lambda^{1/15},\quad \langle \Phi_{1,3}\rangle=2.00314\ldots\ \lambda^{8/15}.
\end{equation}
These values are in good agreement with numerical calculations~\cite{GM} and have been verified for the operator $\Phi_{1,3}$ with a higher precision in~\cite{T1}. As a result we show that the proposed construction of the free-field representation can be used for the multi-particle form factors computation in the IMMF.

\section{Conclusion}
Let us summarize the results. In the framework of the free-field representation we construct a space of solutions to the form factor axioms in the Bullough-Dodd model. The proposed free-field representation differs from the conventional by excluding the minimal two-particle form factors from the construction. As a consequence, the calculation of the multi-particle form factors reduces to the computation of particular matrix elements. The $N$-particle matrix element is given by the sum of $3^N$ terms whose calculation can be performed by using Wick's averaging procedure generating rational functions in the variables $x_i=e^{\theta_i}$ only. The particular terms in the sum may possess non-physical poles. However, we consider the operator algebra in detail and prove that the contribution of these poles in the whole sum vanishes. 

A simple analytic structure of the proposed matrix elements allows us to obtain explicit recurrence relations between them. By using these relations we prove that the form factors of the exponential operators are consistent with the quantum equation of motion and satisfy reflection properties. Besides, in the framework of the free-field representation we compute the form factors of the exponential operators up to four-particle one.

We consider the quantum group restrictions of the imaginary coupling Bullough-Dodd model. In contrast to Lukyanov's construction of the free-field representation where additional poles necessarily arises from the two-particle minimal form factors contribution, the proposed construction can be easily used to obtain the free-field representation for the lightest particle form factors in the $\Phi_{1,2}$ perturbed minimal models. Since we exclude the minimal form factors from the construction, the computation of the matrix elements is performed by Wick's averaging procedure generating the rational functions with the same set of poles as in the Bullough-Dodd model.  

A particular example is provided by the Ising model in a magnetic field. It is a challenging problem to obtain the free-field representations associated with the $E_8$ algebra directly. The remarkable connection between this model and the Bullough-Dodd model was established in~\cite{Shiraishi,BL}. In this paper we present the free-field representation for the lightest-particle form factors in the IMMF in convenient form. As a consistency check we verify that the results obtained in the framework of the free-field representation are in full agreement with those ones obtained by solving the bootstrap equations.

\section*{Acknowledgments}
I am grateful to Jun'ichi Shiraishi, Michael Lashkevich and Yaroslav Pugai for helpful discussions. I would like to thank G\'abor Tak\'acs and Gesualdo Delfino for reading the first version of the paper and useful remarks. The work was supported by the Federal program ``Scientific and Scientific-Pedagogical Personnel of Innovational Russia'' on 2009-2013 (state contracts No. P1339 and No. 02.740.11.5165) and by the Russian Ministry of Science and Technology under the Scientific Schools grant 6501.2010.2.

\Appendix

\section{Analytic properties of the function $J_{N,a}(x_1,\ldots,x_N)$}\label{BDA}
Consider the function $J_{N+1,a}(z,x_1,\ldots,x_N)$, which is a symmetric function in the variables $z,x_1,\ldots,x_N$. Therefore it is sufficient to determine its analytic properties with respect to the variable $z$ alone considering the other ones as a parameter. Due to the form factor axioms~\eqref{ST:kpole} and~\eqref{ST:bpole} one can expect that the only poles of the function $J_{N+1,a}(z,x_1,\ldots,x_N)$ are those at the points
\begin{equation}\label{BDA:poles}
z=x_i\omega^{\pm2},\qquad z=-x_i,
\end{equation}
where the former poles correspond to the rapidity difference $\theta=2\pi i/3$ and, therefore, to the bound state poles, while the later one gives $\theta=i\pi$ and corresponds to the kinematical pole. Let us prove that the matrix element
\begin{equation}\label{BDA:ME}
\langle t(z)t(x_1)\ldots t(x_N)\rangle
\end{equation}
being considered as a function in the variable $z$ has simple poles at~\eqref{BDA:poles} as well. Besides we determine the corresponding residues explicitly. We consider the operator product expansion
\begin{equation}\label{BDA:tt}
t(z)t(x).
\end{equation}
By using~\eqref{BD:ll} one can chow that its simple poles are those at the points
\begin{equation}\label{BD:ttpoles}
	z=x\omega^{\pm1},\quad z=x\omega^{\pm2},\quad z=-x.
\end{equation}
Let us consider these poles consequentially.
\begin{enumerate}
\item The poles at $z=x\omega^{\pm1}$. As it follows from~\eqref{BDA:poles} these poles correspond nor dynamical nor kinematical ones. Let us prove that the corresponding residues of the matrix element~\eqref{BDA:ME} vanish. The singular part of the operator product expansion~\eqref{BDA:tt} is given by
\begin{equation}\label{BD:ResPole1}
	(z-x\omega^{\pm1})t(z)t(x)\Bigr|_{z=x\omega^{\pm1}}=\pm x\omega^{\pm1}  r   h ^2\left(:\lambda^0(x\omega^{\pm1})\lambda^0(x):-:\lambda^\mp(x\omega^{\pm1})\lambda^\pm(x):\right).
\end{equation}
where we introduce the following notation,
\begin{equation}\label{BD:fres}
	\Res_{z=x\omega^{\pm1}}f\left( \frac{z}{x} \right)=\pm x\omega^{\pm1} r ,\qquad  r =\frac{h^2-1}{\omega-\omega^{-1}}.
\end{equation}
The operators that appears in the right hand side of~\eqref{BD:ResPole1} satisfy the following relations
\begin{equation}\label{BD:op1}
	\langle:\lambda^0(x\omega^{\pm1})\lambda^0(x):\lambda^\sigma(x')\rangle=\langle:\lambda^\mp(x\omega^{\pm1})\lambda^\pm(x):\lambda^\sigma(x')\rangle,\quad\sigma=\pm,0.
\end{equation}
Therefore, from~\eqref{BD:ResPole1} together with~\eqref{BD:op1} we obtain
\begin{equation}
	(z-x_n\omega^{\pm1})\langle t(z)t(x_1)\ldots t(x_n)\ldots t(x_N)\rangle\Bigr|_{z=x_n\omega^{\pm1}}=0.
\end{equation}
\item The dynamical poles at $z=x\omega^{\pm2}$. At these poles the singular part of the operator product expansion is given by
\begin{equation}\label{BD:ResPole2}
	(z-x\omega^{\pm2})t(z)t(x)\Bigr|_{z=x\omega^{\pm2}}= \pm x\omega^{\pm2} h  r \tilde t(x\omega^{\pm1}),
\end{equation}
where we introduce the operator $\tilde t(x)$,
\begin{equation}\label{BD:ttilde}
	\tilde t(x)=\gamma :\lambda^0(x\omega^{+1})\lambda^+(x\omega^{-1}) :+\gamma^{-1}:\lambda^-(x\omega^{+1})\lambda^0(x\omega^{-1}) : +  h  :\lambda^-(x\omega^{+1})\lambda^+(x\omega^{-1}) :.
\end{equation}
Notice the similarity between this operator and those one defined  in~\eqref{BD:t}. In every matrix element containing the operator $\tilde t(z)$ one can replace it by $t(z)$ taking into account that 
\begin{equation}\label{BD:op2}
	\langle \tilde t(z)t(x_1)\dots t(x_N) \rangle=\prod_{n=1}^Nf\Bigl( \frac{z}{x_n} \Bigr) \langle  t(z)t(x_n)\dots t(x_N) \rangle.
\end{equation}
Therefore from~\eqref{BD:ResPole2} and~\eqref{BD:op2} we obtain
\begin{multline}\label{BD:Res2}
	(z-x_n\omega^{\pm2})\langle t(z)t(x_1)\ldots t(x_n)\ldots t(x_N)\rangle\Bigr|_{z=x_n\omega^{\pm2}} =\\
	=x_n\omega^{\pm2}B^\pm_n(x_1,\ldots,x_N)\langle t(x_n\omega^{\pm1})t(x_1)\ldots\cancel {t(x_n)}\ldots,t(x_N)\rangle,
\end{multline}
where
\begin{equation}\label{BD:B}
B_n^{\pm}(x_1,\ldots, x_N)= h   r \prod_{i\neq n} f\Bigl( \frac{x_n}{x_i}\omega^{\pm1} \Bigr).
\end{equation}
\item The kinematical pole at $z=-x$. The singular part of the operator product expansion at this pole is the following
\begin{equation}\label{BD:ResPole3}
	(z+x)t(z)t(x)\Bigr|_{z=-x}=x f(\omega^2) r  \left(s(x\omega^{-3/2})-s(x\omega^{3/2})\right)
\end{equation}
where the operator $s(x)$ is given by
\begin{equation}
	s(x)=: \lambda^+(x\omega^{-3/2})\lambda^-(x\omega^{3/2}):\ .
\end{equation}
Being inserted into the matrix element this operator simply produces an overall factor,
\begin{equation}\label{BD:op3}
	\langle  s(z)t(x_1)\dots t(x_N) \rangle =\prod_{n=1}^N f \Bigl( \frac{z}{x}\omega^{1/2} \Bigr)f\Bigl( \frac{z}{x}\omega^{-1/2} \Bigr) \langle  t_1(x_1)\dots t(x_N) \rangle
\end{equation}
Consequently, from~\eqref{BD:ResPole3} and~\eqref{BD:op3} we get
\begin{multline}\label{BD:Res3}
	(z+x_n)\langle t(z)t(x_1)\ldots t(x_n)\ldots t(x_N)\rangle\Bigr|_{z=-x} =\\
	=x_n K_n(x_1,\ldots,x_N)\langle t(x_1)\ldots\cancel {t(x_n)}\ldots,t(x_N)\rangle,
\end{multline}
where
\begin{equation}\label{BD:K}
	\begin{aligned}
		K_n(x_1,\ldots,x_N)=&f(\omega^2)  r \biggl(\prod_{i\neq n}f\Bigl(\frac{x_n}{x_i}\omega^{-2}\Bigr)f\Bigl(\frac{x_n}{x_i}\omega^{-1}\Bigr)-\prod_{i\neq n}f\Bigl(\frac{x_n}{x_i}\omega^{2}\Bigr)f\Bigl(\frac{x_n}{x_i}\omega\Bigr)\biggr).
	\end{aligned}
\end{equation}
\end{enumerate}
Thus we prove that the matrix element~\eqref{BDA:ME} as a function in the variable $z$ has simple poles at~\eqref{BDA:poles} and the corresponding residues~\eqref{BD:Res2} and~\eqref{BD:Res3} can be represented as a matrix elements of the less number but the same operators $t(x_i)$.

\section{The regular part of the function $J_{N,a}(x_1,\ldots,x_N)$}\label{BDB}
Consider the function $J_{N,a}(z,X)$ as an analytic function of the variable $z$ depending on the parameter $X=\{x_1,\ldots,x_N\}$. It is possible to separate the contribution of the poles from the regular part,
\begin{equation}\label{BD:rec}
	\begin{aligned}
		J_{N+1,a}(z,X)=\ &J^{(\infty)}_{N+1,a}(z,X)\\
		&+\sum\limits_{n=1}^N\frac{x_n}{z+x_n}K_n(X)J_{N-1}(\hat X_n)\\
		&+\sum\limits_{n=1}^N\frac{x_n\omega^{2}}{z-x_n\omega^2}B_n^+(X)J_{N,a}(x_n\omega,\hat X_n)\\
		&-\sum_{n=1}^N\frac{x_n\omega^{-2}}{z-x_n\omega^{-2}}B_n^-(X)J_{N,a}(x_n\omega^{-1},\hat X_n).
	\end{aligned}
\end{equation}
where we introduce the notation $\hat X_n=\{x_1,\ldots,\cancel{x}_n,\ldots,x_N\}$.
The function $J^{(\infty)}_{N+1,a}(z,X)$ is a regular function in the variable $z$ everywhere except the points $z=0$ and $z=\infty$. Since the sum over poles is of order $O(z^{-1})$ as $z\to\infty$, the asymptotic behavior of the function $J_{N+1,a}(z,X)$ as a function in the variable $z$ is governed by $J^{(\infty)}_{N+1,a}(z,X)$. The expansion~\eqref{BD:rec} is applicable in the vicinity of zero. 

One can consider the expansion of the function $J_{N+1,a}(z,X)$ in the vicinity of the point $z=\infty$,
\begin{equation}\label{BD:rec0}
	\begin{aligned}
	J_{N+1,a}(z,X)=\ &J^{(0)}_{N+1,a}(z,X)\\
	&-\sum\limits_{n=1}^N\frac{x_n^{-1}}{z^{-1}+x_n^{-1}}K_n(X) J_{N-1}(\hat{X}_n)\\
	&-\sum\limits_{n=1}^N\frac{x_n^{-1}\omega^{2}}{z^{-1}-x_n^{-1}\omega^2}B_n^+(X)J_{N,a}(x_n\omega^{+1};\hat{X}_n)\\
	&+\sum_{n=1}^N\frac{x_n^{-1}\omega^{-2}}{z^{-1}-x_n^{-1}\omega^{-2}}B_n^-(X)J_{N,a}(x_n\omega^{-1};\hat{X}_n),
	\end{aligned}
\end{equation}
where the function $J^{(0)}_{N+1,a}(z,X)$ is regular everywhere except the points $z=0,\infty$. At the vicinity of the point $z=0$ the behavior of the function $J_{N+1,a}(z,X)$ is governed by $J^{(0)}_{N+1,a}(z,X)$. 

Notice that the expansions~\eqref{BD:rec} and~\eqref{BD:rec} are applicable for any kind of operators under consideration, i.e.\ not only for exponential but for descendant operators as well. At this point we restrict our attention to the exponential operators only. As discussed in section~\ref{ST} the form factors of the exponential operators should satisfy the factorization property. Consequently the only unknown functions $J^\infty_{N+1,a}(z,X)$ and $J^{(0)}_{N+1,a}(z,X)$ in the recurrence relations can be fixed easily. Indeed, consider the asymptotic of the function $J_{N+1,a}(z,X)$ as $z\to0$ and $z\to\infty$. Using the factorization property~\eqref{ST:factor} we get
\begin{equation}\label{BD:Jlim}
	J^{(\infty)}_{N+1,a}(z,X)=J^{(0)}_{N+1,a}(z,X)=J_{1,a}J_{N,a}(X),
\end{equation}
where the one-particle form factor is given by 
\begin{equation}
	J_{1,a}=\gamma+\gamma^{-1}+ h .
\end{equation}
Taking into account explicit expressions for $ h $ and $\gamma$ we obtain the proposition~\ref{BD:prop}.

In the remain part of this section we obtain the useful identity for the functions $J_{N,a}(x_1,\ldots,x_N)$. Let us define the function
\begin{multline}\label{BD:D}
	D_{n,a}(X)=\sum\limits_{n=1}^N K_n(X)J_{N-1}(\hat{X}_n)+ \sum\limits_{n=1}^N x_n\omega^{2}B_n^+(X)J_{N,a}(x_n\omega^{+1};\hat{X}_n)-\\
	-\sum\limits_{n=1}^N x_n\omega^{-2}B_n^-(X)J_{N,a}(x_n\omega^{-1};\hat{X}_n).
\end{multline}
From the expansions~\eqref{BD:rec} and~\eqref{BD:rec0} it follows that
\begin{equation}\label{BD:J-J}
	J^{(0)}_{N+1,a}(z,X)-J^{(\infty)}_{N+1,a}(z,X)=D_{N,a}(X)
\end{equation}
Taking into account~\eqref{BD:Jlim} we get
\begin{equation}\label{BD:D0}
	D_{N,a}=0.
\end{equation}
This equality provides a non trivial identity for the functions $J_{N,a}(x_1,\ldots,x_N)$ and we use it in Appendix~\ref{EM} to prove that the form factors of the exponential operators are compatible with the quantum equation of motion.

\section{The equation of motion for form factors}\label{EM}
In this section we prove that form factors in the Bullough-Dodd model are consistent with the equation of motion,
\begin{equation}
	\partial\bar\partial\varphi=8\sqrt2\pi\mu b\left(e^{\sqrt2 b\varphi}-e^{-\frac{b}{\sqrt2}\varphi}\right).
\end{equation}
The derivatives of a field produce multiplication of its form factors by the components of the momentum according to the usual rule $P_\mu\to i\partial_\mu$. Let us introduce the notation
\begin{equation}
	S_n(X)=x_1^n+\ldots+x_N^n,\qquad S_n(z,X)=z^n+S_n(X),
\end{equation}
where $X=\{x_1,\ldots,x_N\}$. Let $z=e^\theta$ and $x_n=e^{\theta_n}$. The components of the momentum are given by
\begin{equation}
P_z(\theta,\theta_1,\ldots,\theta_N)=-\frac{m}{2}S_1(z,X),\quad P_{\bar z}(\theta,\theta_1,\ldots,\theta_N)=\frac{m}{2}S_{-1}(z,X).
\end{equation}
Therefore,
\begin{equation}
	\langle vac|\partial\bar\partial\varphi|\theta,\theta_1,\ldots,\theta_N\rangle=-\frac{m^2}{4}S_1(z,X)S_{-1}(z,X)\frac{d}{da}f_a(\theta,\theta_1,\ldots,\theta_N)\Bigr|_{a=0}.
\end{equation}
Let
\begin{equation}
	J'_{N+1}(z,X)=\frac{d}{da}J_{N+1,a}(z,X)\Bigr|_{a=0}.
\end{equation}
Using these notation we can represent the equation of motion in the following form
\begin{equation}\label{EM:JEM}
	S_1(z,X)S_{-1}(z,X)J'_{N+1}(z,X)=A\Bigl(J_{N+1,\sqrt2b}-J_{N+1,-b/\sqrt2}\Bigr),
\end{equation}
where the constant $A$ is given by
\begin{equation}
	A=-\frac{8\sqrt2\pi\mu b}{m^2}\langle e^{\sqrt2b\varphi}\rangle,
\end{equation}
and we take into account that the vacuum expectation values of the corresponding exponential operators coincide~\cite{FLZZ}
\begin{equation}
\langle e^{\sqrt2b\varphi}\rangle = \langle e^{-b\varphi/\sqrt2}\rangle = \frac{m^2}{\mu}\frac{1}{48\sqrt3 Qb\sin\bigl(\frac{\pi}{6Qb}\bigr)\sin\bigl( \frac{\pi}{3Qb}\bigr)}.
\end{equation}
Let us prove the equation of motion~\eqref{EM:JEM} recursively. Suppose that this equation is valid for some value $N$. Taking derivatives of both sides of the recurrence relation~\eqref{BD:ThR} we get
\begin{equation}\label{EM:SSJ}
	\begin{aligned}
	S_1(z,X)S_{-1}(z,X)&J'_{N+1}(z,X)=\sum_{n=1}^N x_nK_n(\hat X_n)S_{-1}(\hat X_n)J'_{N-1}(\hat X_n)+\\
	&+\sum_{n=1}^N\frac{x_n}{z+x_n}K_n(X)S_1(\hat X_n)S_{-1} (\hat X_n)J'_{N-1}(\hat X_n)+\\
	&+\sum_{n=1}^N x_n \omega^2B^+_n(X)S_{-1}(x_n \omega,\hat X_n)J'_{N+1}(x_n \omega, \hat X_n)+\\
	&+\sum_{n=1}^N\frac{x_n \omega^2}{z-x_n \omega^2}B^+_n(X)S_1(x_n \omega,\hat X_n)S_{-1}(x_n \omega,\hat X_n)J'_N(x \omega,\hat X_n)-\\
	&-\sum_{n=1}^N x_n \omega^{-2}B^-_n(X)S_{-1}(x_n \omega^{-1},\hat X_n)J'_{N+1}(x_n \omega^{-1}, \hat X_n)-\\
	&-\sum_{n=1}^N\frac{x_n \omega^{-2}}{z-x_n \omega^{-2}}B^-_n(X)S_1(x_n \omega^{-1},\hat X_n)S_{-1}(x_n \omega^{-1},\hat X_n)J'_N(x \omega^{-1},\hat X_n).
	\end{aligned}
\end{equation}
Here in the right hand side one can apply the induction hypothesis~\eqref{EM:JEM}. Besides we use the following identities,
\begin{multline}
	\sum_{n=1}^N\frac{x_n}{S_1(\hat X_n)}K_n(X)J_{N-1,a}(\hat X_n)+\sum_{n=1}^N\frac{x_n \omega^2}{S_1(x_n \omega,\hat X_n)}B^+_n(X)J_{N,a}(x_n \omega,\hat X_n)\\
	-\sum_{n=1}^N\frac{x_n \omega^{-2}}{S_1(x_n \omega^{-1},\hat X_n)}B^-_n(X)J_{N,a}(x_n \omega^{-1},\hat X_n)\\
	=-J_{N+1,a}(-S_1(X),X)+J_{1,a}J_{N,a}(X),
\end{multline}
and
\begin{multline}
	\sum_{n=1}^N\frac{x_n}{z+x_n}K_n(X)J_{N-1,a}(\hat X_n)+\sum_{n=1}^N\frac{x_n \omega^2}{z-x_n \omega^2}B^+_n J_{N,a}(x_n \omega,\hat X_n)\\
	-\sum_{n=1}^N\frac{x_n \omega^{-2}}{z-x_n \omega^{-2}}B^-_n(X)J_{N,a}(x_n \omega^{-1},\hat X_n)\\
	=J_{N+1,a}(z,X)-J_{1,a}J_{N,a}(X).
\end{multline}
where we use~\eqref{BD:J-J}. Therefore~\eqref{EM:SSJ} gets the following form
\begin{multline}\label{EM:SSJ2}
	S_1(z,X)S_{-1}(z,X)J'_{N+1}(z,X)=A\Bigl(J_{N+1,\sqrt2b}(z,X)-J_{N+1,b/\sqrt2}(z,X)-\Bigr.\\ \Bigl.-J_{N+1,\sqrt2b}(-S_1(X),X)+J_{N+1,b/\sqrt2}(-S_1(X),X)\Bigr).
\end{multline}
Compare this equation with those of~\eqref{EM:JEM}. To prove the equation of motion for $N+1$ it is sufficient to prove that the last two terms vanishes,
\begin{equation}\label{EM:JS}
	-J_{N+1,\sqrt2b}(-S_1(X),X)+J_{N+1,b/\sqrt2}(-S_1(X),X)=0.
\end{equation}
In fact, these terms vanish separately independently on $a$. To prove~\eqref{EM:JS} notice that the function $J_{N+1,a}(-S_1(X),X)$ is symmetric in the variables $x_i$. Therefore, any of these variables can be chosen for $z$, i.e.
\begin{equation}
	J_{N+1,a}(-S_1,X)=J_{N+1,a}(-S_1(x_{N+1},\hat X_j),x_{N+1},\hat X_j).
\end{equation}
The left hand side is $x_N$-independent, while the right hand side is $x_j$-independent. This proves that the function $J_{N+1,a}(-S_1(X),X)$ is constant in all variables $X$. Let us show that this function is equal to zero. Consider the limit $x_N\to\infty$. From the recurrence relation we obtain
\begin{equation}
	\begin{aligned}
		&J_{N+1,a}(-S_1(X);X)&=\ &J_{1,a}J_{N,a}(X)+\sum\limits_{n=1}^N\frac{x_n}{S_1(\hat X_n)}K_n(X)J_{N-1,a}(\hat{X}_n)\\
		&&&+\sum\limits_{n=1}^N\frac{x_n\omega^{2}}{S_1(x_n\omega,\hat X_n)}B^+_n(X)J_{N,a}(x_n\omega^{+1};\hat{X}_n)\\
		&&&-\sum_{n=1}^N\frac{x_n\omega^{-2}}{S_1(x_n \omega^{-1},\hat X_n)}B^-_n(X)J_{N,a}(x_n\omega^{-1};\hat{X}_n)
	\end{aligned}
\end{equation}
Since the left hand side is a constant, we can calculate it in the limit $x_N\to\infty$. In this limit the only non-vanishing terms in the sums in the right hand side are those with $n=N$. Using the factorization property one can show that in the leading order in $x_N$ the function $J_{N+1,a}(-S_1(X);X)\to0$. Since this function is a constant we prove~\eqref{EM:JS}. Consequently we prove that the form factors $J_{N,a}(x_1,\ldots,x_N)$ indeed satisfy the equation of motion of the Bulloug-Dodd model.

\section{The four-point symmetric polynomial $\Lambda_4(x_1,\ldots,x_4)$}\label{4p}
In this appendix we present coefficients of the polynomial $\Lambda_4(\sigma_1,\sigma_2,\sigma_3,\sigma_4)$. Let us choose the following parametrization,
\begin{equation}\label{A:L4}
	\Lambda_4(\sigma_1,\sigma_2,\sigma_3 \sigma_4)=\sum\limits_{ \substack{i,j,k,l=0\\ i+2j+3k+4l=18\\ i+j+k+l\leq9}}^{18} c_{i,j,k,l} \sigma_1^i\sigma_2^j\sigma_3^k\sigma_4^l.
\end{equation}
In this notation the coefficients $c_{i,j,k,l}$ are the following
\begin{align*}
	c_{0,0,2,3}=&\ -J^4-J  (-1+ h  ^2 )  (3 J-24  h  +7 J^2  h  +4 J  h  ^2+74  h  ^3-6 J^2  h  ^3\\*
&-13 J  h  ^4-85  h  ^5+J^2  h  ^5+7 J  h  ^6+45  h  ^7-J  h  ^8-11  h  ^9+ h  ^{11} )\\
	c_{0,1,0,4}=&\ 0\\ c_{0,1,4,1}=&\ J  h    (4-5  h  ^2+ h  ^4 )  (-3+J^2+J  h  +7  h  ^2-J  h  ^3-5  h  ^4+ h  ^6 )\\
	c_{0,2,2,2}=&\ 2 J^4+J  (-1+ h  ^2 )^2  (-6 J+21  h  -5 J^2  h  -5 J  h  ^2-\\
&-46  h  ^3+J^2  h  ^3+6 J  h  ^4+34  h  ^5-J  h  ^6-10  h  ^7+ h  ^9 )\\
	c_{0,3,0,3}=&\ 0\\ c_{0,3,4,0}=&\ J^4-J^2  (-1+ h  ^2 )  (-3+J  h  +4  h  ^2- h  ^4 )\\
	c_{0,4,2,1}=&\ -J^4+J  (-1+ h  ^2 )  (-3 J-3  h  +2 J^2  h  +5 J  h  ^2+7  h  ^3-2 J  h  ^4-5  h  ^5+ h  ^7 )\\
	c_{0,5,0,2}=&\ 0\\
	c_{1,0,3,2}=&\ 3 J^4+J  (-1+ h  ^2 )  (9 J-24  h  +13 J^2  h  +4 J  h  ^2+74  h  ^3-12 J^2  h  ^3-25 J  h  ^4\\*
&-85  h  ^5+2 J^2  h  ^5+14 J  h  ^6+45  h  ^7-2 J  h  ^8-11  h  ^9+ h  ^{11} )\\
	c_{1,1,1,3}=&\ J^4+J  h    (-1+ h  ^2 )  (-27+7 J^2-40 J  h  +33  h  ^2-6 J^2  h  ^2\\*
&+74 J  h  ^3+46  h  ^4+J^2  h  ^4-44 J  h  ^5-93  h  ^6+11 J  h  ^7+52  h  ^8-J  h  ^9-12  h  ^{10}+ h  ^{12} )\\
	c_{1,1,5,0}=&\ -J^2  h    (J+ h  - h  ^3 )  (4-5  h  ^2+ h  ^4 )\\
	c_{1,2,3,1}=&\ -J^4+J  (-1+ h  ^2 )^2  (3 J+15  h  +J^2  h  +17 J  h  ^2-19  h  ^3-9 J  h  ^4+4  h  ^5+J  h  ^6 )\\
	c_{1,3,1,2}=&\ -2 J^4-J  h    (-1+ h  ^2 )^2  (27-5 J^2+5 J  h  -42  h  ^2+J^2  h  ^2-J  h  ^3+17  h  ^4-2  h  ^6 )\\
	c_{1,4,3,0}=&\ -J^4+J^3  h    (-1+ h  ^2 )\\ c_{1,5,1,1}=&\ J^4-J^2  h    (-1+ h  ^2 )  (2 J+ h  - h  ^3 )\\
	c_{2,0,0,4}=&\ -J^4-J  (-1+ h  ^2 )  (3 J-24  h  +7 J^2  h  +4 J  h  ^2+74  h  ^3-6 J^2  h  ^3\\*
&-13 J  h  ^4-85  h  ^5+J^2  h  ^5+7 J  h  ^6+45  h  ^7-J  h  ^8-11  h  ^9+ h  ^{11} )\\
	c_{2,0,4,1}=&\ -3 J^4-J^2  (-1+ h  ^2 )^2  (-9-5 J  h  -5  h  ^2+J  h  ^3+6  h  ^4- h  ^6 )\\
	c_{2,1,2,2}=&\ -3 J^4-J  h    (-1+ h  ^2 )  (-54+21 J^2-36 J  h  +138  h  ^2-14 J^2  h  ^2+69 J  h  ^3\\*
&-118  h  ^4+2 J^2  h  ^4-43 J  h  ^5+38  h  ^6+11 J  h  ^7-4  h  ^8-J  h  ^9 )\\
	c_{2,2,0,3}=&\ 2 J^4+J  (-1+ h  ^2 )^2  (-6 J+21  h  -5 J^2  h  -5 J  h  ^2-\\
&-46  h  ^3+J^2  h  ^3+6 J  h  ^4+34  h  ^5-J  h  ^6-10  h  ^7+ h  ^9 )\\
	c_{2,2,4,0}=&\ -J^4+J^2  (-1+ h  ^2 )  (-3-4 J  h  +3  h  ^2+J  h  ^3 )\\
	c_{2,3,2,1}=&\ 3 J^4-J  (-1+ h  ^2 )  (-6 J-6  h  -3 J^2  h  +J  h  ^2+13  h  ^3+2 J^2  h  ^3+7 J  h  ^4-8  h  ^5-2 J  h  ^6+ h  ^7 )\\
	c_{2,4,0,2}=&\ -J^4+J  (-1+ h  ^2 )  (-3 J-3  h  +2 J^2  h  +5 J  h  ^2+7  h  ^3-2 J  h  ^4-5  h  ^5+ h  ^7 )\\
	c_{2,5,2,0}=&\ 0\\
	c_{3,0,1,3}=&\ 3 J^4+J  (-1+ h  ^2 )  (9 J-24  h  +13 J^2  h  +4 J  h  ^2+74  h  ^3-12 J^2  h  ^3-25 J  h  ^4\\*
&-85  h  ^5+2 J^2  h  ^5+14 J  h  ^6+45  h  ^7-2 J  h  ^8-11  h  ^9+ h  ^{11} )\\
	c_{3,0,5,0}=&\ J^4-J^2  (-1+ h  ^2 )  (-3+J  h  +4  h  ^2- h  ^4 )\\
	c_{3,1,3,1}=&\ 3 J^4+J  h    (-1+ h  ^2 )  (-3+13 J^2+4 J  h  +7  h  ^2-8 J^2  h  ^2-5 J  h  ^3-5  h  ^4+J^2  h  ^4+J  h  ^5+ h  ^6 )\\
	c_{3,2,1,2}=&\ -J^4+J  (-1+ h  ^2 )^2  (3 J+15  h  +J^2  h  +17 J  h  ^2-19  h  ^3-9 J  h  ^4+4  h  ^5+J  h  ^6 )\\
	c_{3,3,3,0}=&\ J^4\\ c_{3,4,1,1}=&\ -J^4+J^3  h    (-1+ h  ^2 )\\
	c_{4,0,2,2}=&\ -3 J^4-J^2  (-1+ h  ^2 )^2  (-9-5 J  h  -5  h  ^2+J  h  ^3+6  h  ^4- h  ^6 )\\
	c_{4,1,0,3}=&\ J  h    (4-5  h  ^2+ h  ^4 )  (-3+J^2+J  h  +7  h  ^2-J  h  ^3-5  h  ^4+ h  ^6 )\\
	c_{4,1,4,0}=&\ -J^4+J^3  h    (-1+ h  ^2 )\\
	c_{4,2,2,1}=&\ -J^4+J^2  (-1+ h  ^2 )  (-3-4 J  h  +3  h  ^2+J  h  ^3 )\\
	c_{4,3,0,2}=&\ J^4-J^2  (-1+ h  ^2 )  (-3+J  h  +4  h  ^2- h  ^4 )\\
	c_{5,0,3,1}=&\ J^4-J^2  (-1+ h  ^2 )  (-3+J  h  +4  h  ^2- h  ^4 )\\
	c_{5,1,1,2}=&\ -J^2  h    (J+ h  - h  ^3 )  (4-5  h  ^2+ h  ^4 ).
\end{align*}


\begin{thebibliography}{99}
\bibitem{BPZ} 
	A.A. Belavin, A.M. Polyakov and A.B. Zamolodchikov,
	\emph{Nucl. Phys.} \textbf{B241}
	(1984), 333;
\bibitem{ABZIMMF}
	A.B. Zamolodchikov,
	\emph{Advanced Studies in Pure Mathematics} \textbf{19}
	(1989), 641;
\bibitem{ABZIMMF2}
	A.B. Zamolodchikov,
	\emph{Int. J. Mod. Phys} \textbf{A4}
	(1989), 4235;
\bibitem{T}
	G. Takacs,
	\emph{Nucl.Phys.} \textbf{B489}
	(1997), 532;
\bibitem{RS} N. Reshetikhin and F. Smirnov,
	\emph{Comm. Math. Phys.} \textbf{131}
	(1990), 157;
\bibitem{BerLe}
	D. Bernard and A. LeClair, 
	\emph{ Nucl. Phys.} \textbf{B340} 
	(1990), 721;
\bibitem{Smirnov}
	F.A. Smirnov,
	\emph{Int. J. Mod. Phys.} \textbf{A6}
	(1991), 1407;
\bibitem{Eft}
	C. J. Efthimiou, 
	\emph{Nucl. Phys.} \textbf{B398}
	(1993), 697;	
\bibitem{BKW} 
	F.A. Smirnov,
	\emph{Form factors in completely integrable models of quantum field theories}, World Scientific,
	(1992);
\bibitem{CM}
	J. Cardy and G. Mussardo,
	\emph{Nucl. Phys.} \textbf{B340}
	(1990), 387;
\bibitem{DMS}
	G. Delfino, G. Mussardo and P. Simonetti,
	\emph{Nucl. Phys.} \textbf{B737}
	(1996), 469;
\bibitem{ZZ}
	A. Zamolodchikov and I. Ziyatdinov,
	\emph{Nucl.Phys.} \textbf{B849}
	(2011), 654;
\bibitem{DelfinoM}
	G. Delfino, P. Grinza and G. Mussardo,
	\emph{Nucl.Phys.} \textbf{B737}
	(2006), 291;
\bibitem{T1}
	B. Pozsgay and G. Takacs,
	\emph{Nucl.Phys.} \textbf{B788},
	(2008), 167;
\bibitem{T2}
	B. Pozsgay and G. Takacs,
	\emph{Nucl.Phys.} \textbf{B788}
	(2008), 209;
\bibitem{Mussardo}
	A. Koubek and G. Mussardo,
	\emph{Phys.Lett.} \textbf{B311}
	(1993), 193;
\bibitem{Mussardo2}
	A. Fring, G. Mussardo and P. Simonetti,
	\emph{Nucl.Phys.} \textbf{B393}
	(1993), 413;
\bibitem{BD}
	R.K. Dodd and R.K. Bullough,
	\emph{Proc. R. Soc. London} \textbf{A352}
	(1977), 481;
\bibitem{FMS}
	A. Fring, A. Mussardo and P. Simonetti,
	\emph{Phys. Lett.} \textbf{B307}
	(1993), 389;
\bibitem{A}
	C. Acerbi,
	\emph{Nucl.Phys.} \textbf{B497}
	(1997), 589;
\bibitem{Lukyanov}
	S.L. Lukyanov,
	\emph{Commun. Math. Phys.} \textbf{167}
	(1995), 183; 
\bibitem{Lukyanov1}
	S.L. Lukyanov,
	\emph{Mod.Phys.Lett.} \textbf{A12}
	(1997), 2543;
\bibitem{Lukyanov2}
	S.L. Lukyanov,
	\emph{Phys.Lett.} \textbf{B408}
	(1997) 192;
\bibitem{FateevL}
	V.A. Fateev and M. Lashkevich,
	\emph{Nucl.Phys.} \textbf{B696}
	(2004) 301;
\bibitem{FateevPP}
	V.A. Fateev, V.V. Postnikov and Y.P. Pugai,
	\emph{JETP Lett.} \textbf{83}
	(2006) 172;
\bibitem{FateevP}
	V.A. Fateev and Y.P. Pugai,
	\emph{J. Phys.} \textbf{A42}
	(2009), 304013;
\bibitem{Shiraishi}
	Y. Hara, M. Jimbo, H. Konno, S. Odake and J. Shiraishi,
	\emph{arXiv:math/9902150v1 [math.QA]},
	(1999);
\bibitem{BL}
	V.A. Brazhnikov and S.L. Lukyanov,
	\emph{Nucl. Phys.} \textbf{B512}
	(1998), 616;
\bibitem{FL}
	B. Feigin and M. Lashkevich,
	\emph{J.Phys.} \textbf{A42}
	(2009), 304014;
\bibitem{AL}
	O. Alekseev and M. Lashkevich
	\emph{J. High Energy Phys.} \textbf{1007}
	(2010), 095;
\bibitem{FLZZ}
	V. Fateev, S. Lukyanov, A. Zamolodchikov and Al. Zamolodchikov,
	\emph{Nucl.Phys.} \textbf{B516}
	(1998), 652;
\bibitem{Koubek}
	A. Koubek,
	\emph{Int. J. Mod. Phys.} \textbf{A9}
	(1994), 1909;
\bibitem{DM1}
	G. Delfino and G. Mussardo,
	\emph{Nucl. Phys.} \textbf{B455}
	(1995), 724;
\bibitem{DSC}
	G. Delfino, P. Simonetti and J.L. Cardy,
	\emph{Phys. Lett.} \textbf{B387}
	(1996), 327;
\bibitem{MS} G. Mussardo and P. Simonetti,
	\emph{Int. J. Mod. Phys.} \textbf{A9}
	(1994), 3307;
\bibitem{FateevB324}
	V.A. Fateev,
	\emph{Phys. Lett.} \textbf{B324}
	(1994), 45;
\bibitem{GM}
	R. Guida and N. Magnoli
	\emph{Phys.Lett.} \textbf{B411}
	(1997), 127
\end{thebibliography}
\end{document}